\documentclass[aps,prl,twocolumn,showpacs]{revtex4}
\usepackage{amsfonts}
\usepackage[T2A]{fontenc}
\usepackage[cp1251]{inputenc}
\usepackage{amsmath}
\usepackage{amssymb}
\usepackage[english]{babel}
\usepackage{graphicx}

\begin{document}

\title{Ring-shaped spatial pattern of exciton luminescence formed due to the
hot carrier transport in a locally photoexcited electron-hole bilayer}

\author{A. V. Paraskevov$^{1,2}$}

\affiliation{$^1$National Research Center "Kurchatov Institute", Kurchatov Sq. 1, Moscow 123182, Russia\\
$^2$Department of Physics, Loughborough University, Loughborough
LE11 3TU, United Kingdom}

\begin{abstract}
A consistent explanation of the formation of a ring-shaped pattern
of exciton luminescence in GaAs/AlGaAs double quantum wells is
suggested. The pattern consists of two concentric rings around the
laser excitation spot. It is shown that the luminescence rings
appear due to the in-layer transport of hot charge carriers at high
photoexcitation intensity. Interestingly, one of two causes of this
transport might involve self-organized criticality (SOC) that would
be the first case of the SOC observation in semiconductor physics.
We test this cause in a many-body numerical model by performing
extensive molecular dynamics simulations. The results show good
agreement with experiments. Moreover, the simulations have enabled
us to identify the particular kinetic processes underlying the
formation of each of these two luminescence rings.
\end{abstract}


\maketitle

\newpage

\section{Introduction}

Non-equilibrium collective effects in the exciton and
exciton-polariton systems in semiconductor heterostructures are a
subject of intensive studies \cite{Mos, KK, 1990, 1994, Loz-Ber,
But_ring, snoke_nat, L1, L2, Snoke2003, MacDonald, Balats, BLS,
Rap04, Levitov, Den, Ha, Sug06, Szym, Tim, PRB07, stern, polar_nat,
But_new, PRB10, Deng}. A particular attention has been focused on
the beautiful phenomenon discovered experimentally in the system of
interwell excitons in GaAs/AlGaAs double quantum wells (QWs)
\cite{But_ring}: at sufficiently high excitation intensity, a local
photoexcitation of electrons (e) and holes (h) above the exciton
resonance gives rise to a macroscopic ring-shaped pattern of spatial
distribution of the exciton luminescence. The radius of the pattern can
be varied in a wide range by tuning external parameters such as
excitation intensity or gate voltage. Remarkably, at sub-Kelvin
lattice temperatures the external ring of the stationary pattern exhibits a sharp
fragmentation, which could be the signature of a non-equilibrium
macroscopic quantum effect.

To understand the nature of the ring-shaped pattern, one should
build a many-body model that captures local generation of
electron-hole pairs and their spatial dynamics accompanied by the
processes of formation and recombination of excitons. If the exciton
lifetime is long enough, one should also consider the spatial
dynamics of the excitons.

The first theoretical explanation which met these requirements was
based \cite{BLS,Rap04,Den,Ha,But_new} on the diffusive transport model (DTM)
applied to the locally photogenerated holes and equilibrium
electrons which initially were uniformly distributed in the
quantum-well plane. The overlapping region for the hole and electron
spatial distributions apparently gave rise to the ring of exciton
luminescence.

However, this explanation has a lot of evident shortcomings
\cite{com}. For example, if one adds the same number of
photogenerated electrons in the model (in principle, they must be
added to maintain the electroneutrality), then the ring of exciton
luminescence can disappear due to the exciton formation term, which
is simply proportional to the product of the electron and hole
densities. (If these densities decrease monotonically from the
excitation spot center, the luminescence intensity would apparently
follow them.) More specific shortcomings can be found in Appendix A.

In this paper a novel consistent explanation of the ring-shaped pattern
formation is given. The main idea is that there appears an essential
in-plane electric field in the excitation spot region at high enough
excitation power. This field strongly affects the spatial dynamics
of the photogenerated electrons and holes. (We do not consider any
equilibrium carriers at all.) It is shown that there are
contributions to the electric field from two quite different
physical processes. Essentially, due to the one of these
contributions the ring-shaped pattern formation could be understood
in the paradigm of self-organized criticality (SOC) \cite{Bak}. To
test the contribution, we have performed extensive
molecular-dynamics simulations. They have shown that only this
contribution is quite enough for the detailed qualitative
explanation of the ring-shaped pattern. (However, this paper does
not concern the transition to the SOC regime. The parameters for the
simulations have been chosen to be in this regime from the very
beginning.)

This paper is organized as follows. Further in this Section we
introduce some essential properties of double quantum wells and
interwell exciton formation, some experimental results we focus on,
and the formulation of the problem for the research. In Section 2 we
suggest two qualitative explanations ("scenarios") for the
ring-shaped pattern formation and make the corresponding estimates.
Section 3 describes many-body dynamical model and conditions of the
molecular-dynamics simulations performed to investigate the second
scenario in more details. Section 4 contains the results of the
simulations. Section 5 is conclusion and discussion. Finally,
Section 6 consists of three Appendixes.

\begin{figure}[!t]
\begin{center}
\includegraphics[width=8.78cm]{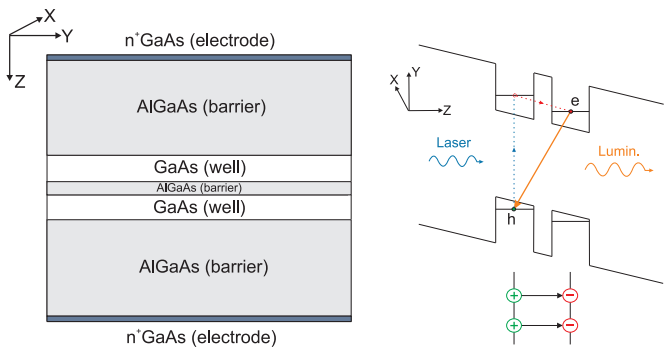}
\caption{\label{fig:fig1} Left: Schematic of a double quantum well
(DQW). The DQW structure is bordered by highly-doped GaAs layers
that serve as external electrodes forming a plane capacitor. In
experiments \protect\cite{But_ring, BLS, But_new} widths of the
layers between the electrodes are $(200$ $nm)(8$ $nm)(4$ $nm)(8$
$nm)(200$ $nm)$, respectively. Right: (Top) Schematic of (i) the DWQ
energy profile along the "growth" axis (Z axis) when there is a
voltage applied between the external electrodes (it results in a
linear bias of the profile) and (ii) interwell exciton formation
(arrows show the path of a photoexcited electron). (Bottom)
Interwell excitons as co-directed classical dipoles.}
\end{center}
\end{figure}

The structure of double QWs used in the experiments \cite{But_ring,
BLS, But_new} is shown in Fig. \ref{fig:fig1}(left). The electron
band-gap energy $E_{gap}$ of the "barrier" (B) layers is larger than
$E_{gap}$ of the "well" (W) layers (Fig. \ref{fig:fig1}(right)) so
that GaAs layers form two rectangular potential wells with the depth
$U_{QW}=\left( E_{gap}\left( B\right) -E_{gap}\left( W\right)
\right) /2\approx0.4$ $eV$.
At a moderate occupation of the wells (i.e., when the number of
electrons in the GaAs conduction band is not macroscopically large;
see next Section for details), a voltage applied to the external
electrodes provides a constant tilt of the DQW potential profile
(Fig. \ref{fig:fig1}(right)). This "gate" voltage, $V_{g}$, is
needed to separate electrons and holes in the different wells
facilitating the formation of interwell excitons. The crossover
between the interwell and intrawell exciton "ground" states takes
place at $V_{g}\approx0.3$ $V$ \cite{But_rev}. (Since in the
experiments \cite{But_ring, BLS, But_new} $V_{g}>0.3$ $V$, the
intrawell excitons are not discussed further.) The stationary laser
pumping comes along $Z$ axis and is used for the formation of a
macroscopic number of photogenerated electron-hole pairs. In
experiments \cite{But_ring, BLS, But_new} the typical laser power
was several hundreds of $\mu W$ and was focused in a spatial spot of
few tens of $\mu m$. The pumping energy was well above the exciton
resonances so that an electron was photoexcited to a high-energy
level of the QW near the continuum. Due to the applied gate voltage,
the electron can then tunnel to the corresponding QW. (The effective
mass of an electron in GaAs is seven times smaller than the one of a
heavy hole so the tunneling is essentially more probable for
electrons than for holes. Note that heavy holes have lower energy
than the light ones in GaAs.) The tunneling might also be
facilitated by the voltage-induced triangular profile of the
barrier. Finally, after a spatiotemporal in-layer relaxation of the
charge carriers, the interwell excitons are formed; they live some
finite time and then annihilate giving rise to the photoluminescence
(PL) pattern in the QW (X-Y) plane.

\begin{figure}[!t]
\begin{center}
\includegraphics[width=8.75cm]{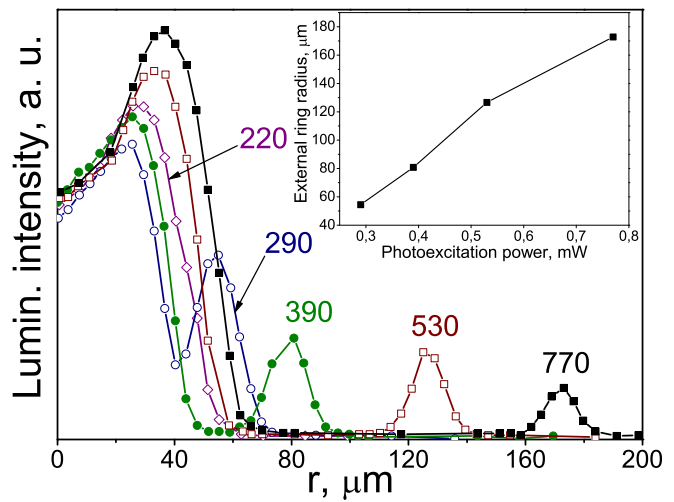}
\caption{\label{fig:fig2} Experimental curves taken from
Ref.\protect\cite{But_ring}: luminescence intensity of interwell
excitons vs distance $r$ from the excitation spot center at
different excitation powers (numbers near curves, in $\mu W$).
Excitation spot radius is about 20 $\mu m$. Inset: Dependence of the
external ring radius on excitation power.}
\end{center}
\end{figure}

Let us now turn to the experimental results that we would like to
explain (Fig. \ref{fig:fig2}).
At small excitation power the PL spatial profile practically follows
the excitation spot (see details in \cite{But_ring}). When
excitation power exceeds some value a thick ring of luminescence
appears near the edge of the excitation spot. In Ref.\cite{But_ring}
it was already seen at excitation power 220 $\mu W$. Hereafter we
call this ring as "internal ring". Finally, when the excitation
power exceeds another critical value (Fig. ~\ref{fig:fig2}), a thin
"external" ring of luminescence appears around the excitation spot
and the inner ring. Everywhere in this paper the words "ring-shaped
pattern" mean these \textit{two concentric rings}.

The formation mechanism of the ring-shaped luminescence pattern is a
subject of the research described below. In particular, we pose the
following questions. (i) Why does the ring-shaped pattern appear
only when the laser excitation power exceeds some critical values?
(ii) What are the kinetic processes which underlie the formation of
the internal and external rings? (iii) Why does the external ring
radius depend strongly on the static gate voltage $V_{g}$
\cite{cond-mat}? (iv) Why the luminescence of intrawell excitons
(i.e., purely "two-dimensional" excitons) does not exhibit the
ring-shaped pattern \cite{But_ring}?

\section{Two scenarios of the ring-shaped luminescence pattern formation}

In general, we believe that the ring-shaped PL pattern appears due
to the transport of hot uncoupled electrons and holes from the
excitation spot at high enough excitation power. During the spatial
spread the carriers relax in kinetic energy emitting phonons and
can eventually form excitons relatively far away from the excitation
spot ("far away" in comparison with the spot radius). We suppose
that the hot charge carriers are formed due to (i) in-layer electric
fields that occur at high pumping power in the excitation spot
region and (ii) high mobilities of the charge carriers in GaAs. We
suggest two particular mechanisms of the electric field occurrence;
they are described in details in next subsections.

\subsection{First scenario: hot carrier transport induced by the external
gate voltage}

\begin{figure}[!th]
\begin{center}
\includegraphics[width=8.8cm]{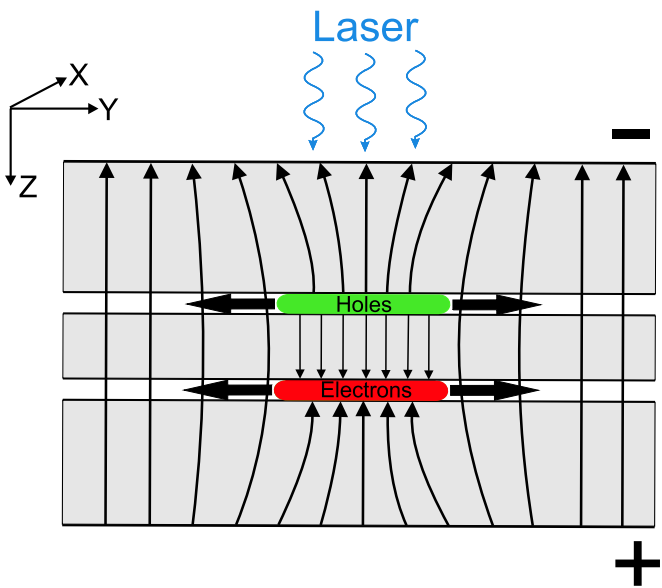}
\caption{\label{fig:fig3} Schematic of the electric field distribution
along the growth axis (Z axis) of DQW structure in the vicinity of laser
excitation spot at high excitation power. (It is supposed that the excitation
spot is far away from the DQW edges and external contacts.) In the spot region (green and red ovals
in the center) the static electric field induced by external gate voltage is
curved due to the presence of a macroscopically large number of photogenerated
charge carriers. The horizontal projections of the field cause in-layer
transport (shown by thick arrows) of the carriers from the excitation spot.}
\end{center}
\end{figure}

The first scenario concerns the screening of the gate voltage
$V_{g}$ by photogenerated carriers in the excitation spot (Fig.
\ref{fig:fig3}) at high excitation power.

The electric induction in the bilayer volume at the laser excitation
spot is $\mathbf{D}=\mathbf{E}+4\pi\mathbf{P}$. Here
$\mathbf{E}=E_{0}\mathbf{e}_{z}$ is the static uniform electric
field generated by $V_{g}$ and $\mathbf{P}$ is polarization of the
medium. If $n$ is the density of electron-hole pairs in the
excitation spot of area $S$ and $d$ is the average distance between
the pairs in different wells (i.e., $d$ is the dipole length), then
$\mathbf{P}=-\left(  nS\right)  \left(  ed\right)  /\left( Sd\right)
\mathbf{e}_{z}=-\left(  ne\right)  \mathbf{e}_{z}$ and
$\mathbf{D}=\left(  E_{0}-4\pi ne\right)  \mathbf{e}_{z}$. In the
experiments \cite{But_rev} the DQW structure was considered as an
insulator. It means that condition
\begin{equation}
E_{0}\gg4\pi en, \label{E00}%
\end{equation}
should be fulfilled in the excitation spot region. In this case the
gate voltage results in a linear slope of the DQW potential energy
profile along $Z$ axis on the value $\delta U\left(  z\right)
\approx-eE_{0}z$ (Fig. \ref{fig:fig1}(right)).

However, at typical value $V_{g}\approx1$ $V$ and
$n\sim10^{10}cm^{-2}$ \cite{But_rev} one gets $eE_{0}\approx
eV_{g}/(2L)\sim10^{4}$ $eV/cm$, where $L=200$ nm is the width of
external barrier of the DQW structure (Fig. \ref{fig:fig1}(left)),
and $4\pi ne^{2}\sim10^{4}$ $eV/cm$. So at expected densities $n\sim
10^{11}\div10^{12}$ cm$^{-2}$ in the excitation spot at high pumping
power, the condition (\ref{E00}) is not fulfilled there. It means
that $Z$-axis component of the resulting field $E_{z}$ is
essentially dependent on $z$ in the excitation spot region. More
importantly, in this case there exists an in-plane component $E_{r}$
of the electric field that will push both electrons in one layer and
holes in another layer away from the excitation spot (Fig.
\ref{fig:fig3}).

\subsection{Second scenario: hot carrier transport induced by the repulsive
in-layer interaction}

If the photogenerated electrons and holes do not leave the
excitation spot for any reason then the higher the pumping power
$P_{ex}$ the higher the carriers densities in the spot. (Since the
excitation is off-resonant, the value of the exciton formation time
is always larger than the time of energy relaxation of the carriers
to reach the exciton transition.) Due to the bilayer geometry, there
exists a value of excitation power at which the carrier densities in
the spot reach the values when repulsive in-layer Coulomb forces
between carriers become stronger than the attractive interlayer
force. To be more specific, let us make an estimate of the
interaction strength in the excitation spot through the
dimensionless interaction parameter $r_{s}$. At small carrier
densities the interaction in the spot is dipole-dipole one rather
than the Coulomb as in the case of a monolayer. In particular, for
electron (or hole) monolayer
\begin{equation}
r_{s}=(e^{2}/\bar{r})/(\hbar^{2}/(m\bar{r}^{2}))=\bar{r}/a_{B}\sim
1/\sqrt{na_{B}^{2}}, \label{int1}%
\end{equation}
where $a_{B}=\hbar^{2}/\left(  me^{2}\right)$ is the Bohr radius, $n$ is
the carrier density in the spot, and $\bar{r}\sim n^{-1/2}$ is the
average distance between carriers. (Hereafter we omit dielectric constant
in the formulae unless making numerical estimates.) Thus, for monolayer the increase
of density $n$ leads to the decrease of interaction. However, in the
case of electron-hole (e-h) bilayer at $\bar {r}>d$, where $d$ is
the interlayer distance, the interaction is dipole-dipole one,
$U=e^{2}d^{2}/\bar{r}^{3}$ rather than $e^{2}/\bar{r}$. This leads
to
\begin{equation}
r_{s}=U/(\hbar^{2}/(m\bar{r}^{2}))=d^{2}/\left(  \bar{r}a_{B}\right)
\sim\left(  d/a_{B}\right)  \sqrt{nd^{2}}, \label{int2}%
\end{equation}
where $n<n_{c}\lesssim d^{-2}\sim10^{12}$ $cm^{-2}$ at
$d\sim10^{-6}$ cm \cite{But_rev}. It is seen that in this case the
interaction increases in accord with the densities. At $n=n_{c}$
(that corresponds to some \textit{critical excitation power} $\left(
P_{ex}\right)  _{c}$) the character of interaction is changing: the
repulsive in-layer interaction becomes dominating and, moreover, one
should put $n=0$ effectively in the estimate (\ref{int1}), i.e., the
repulsive interaction becomes huge. It leads to the appearance of
in-layer electric fields ejecting the electrons and holes from the
excitation spot region. Then the e-h densities in the spot grow to
the critical values again and the ejection process recurs. There is
a direct correspondence between this self-organized ejection, which
keeps the critical values of carrier densities in the spot, and the
avalanches in classical sand-pile model of SOC \cite{Bak, Bak-book}.

Since the mobilities of charge carriers in GaAs QWs are very high
(up to $10^{7}$ cm$^{2}$/(Vs) for electrons at sub-Kelvin
temperatures \cite{mob01,mob02,mob,cp}), the initial kinetic
energies of some part of ejected particles can exceed the threshold
of optical phonon emission. Their relaxation in energy is then very
fast so these carriers will likely form excitons not far away from
the excitation spot. In turn, the carriers with kinetic energy less
than optical phonon energy $\hbar\omega_{opt}$ go further. These
carriers relax relatively slow by emitting acoustic phonons. In GaAs
$\hbar\omega_{opt}\approx37$ $meV$ so the velocities of the carriers
contributing to long-distance transport are less than
$v_{\max}\sim10^{7}$ cm/s. (Some more details about carrier energy
relaxation due to phonon emission can be found in Appendix B.)

In addition, there is an essential difference in mobilities, and in
effective masses, for electrons and heavy holes in GaAs
\cite{mob01,mob02,mob}. This can lead to some difference in in-layer
"stream" velocities of electrons and holes. The latter, in turn,
would result in a suppression of the attractive interlayer Coulomb
force (see Appendix C for details) until the velocities become small
enough due to the acoustic phonon emission. In these conditions the
interlayer exciton formation is also suppressed in some region of
distances from the excitation spot. (Recall, for example, that the
classical scattering cross-section for the Coulomb potential, the
Rutherford cross-section, is proportional to $V^{-4}$, where $V$ is
the initial relative velocity at infinite distance.) The suppression
leads naturally to the formation of a luminescence ring that would
define a circumference of the total luminescence pattern.

Thus, the internal ring of luminescence (Fig. \ref{fig:fig2}) might
appear due to the electrons and holes that had emitted optical
phonon(s) and then quickly formed excitons. In turn, the external
luminescence ring can appear due to the carriers that were under the
optical-phonon emission threshold so they needed larger time (and
corresponding distance) to relax emitting acoustic phonons.

In general, we suggest that at high photoexcitation power there
exists an in-plane electric field $E_{r}$ which consists of two
contributions, "gate voltage-induced" and "in-layer
interaction-induced". Due to high mobilities of charge carriers in
GaAs even a moderate value of $E_{r}$ results in high initial
velocities of the carriers directed towards the outside of the
excitation spot. Large relative velocities of the ejected electrons
and holes lead to the suppression of the interlayer Coulomb
attraction between them. The latter results in the suppression of
exciton formation in some domain of distances from the excitation
spot. In turn, the formation dynamics of ring-shaped luminescence
pattern can be divided in three stages: (i) radial acceleration of
carriers in the excitation spot region due to the in-plane component
of the static electric field that appears at relatively high carrier
density in the excitation spot and due to the in-layer Coulomb
repulsion at high pumping power; (ii) slowing down of unbound
carriers due to emission of optical and acoustic phonons and due to
the ambipolar electric field ("Coulomb drag") (iii) the regime of
strong interlayer Coulomb correlations - formation and optical
recombination of interlayer excitons.

In what follows we focus on the second scenario and test it by
molecular dynamics (MD) simulations.

\section{Molecular dynamics simulations: Numerical model}

\begin{figure}[!th]
\begin{center}
\includegraphics[width=8.8cm]{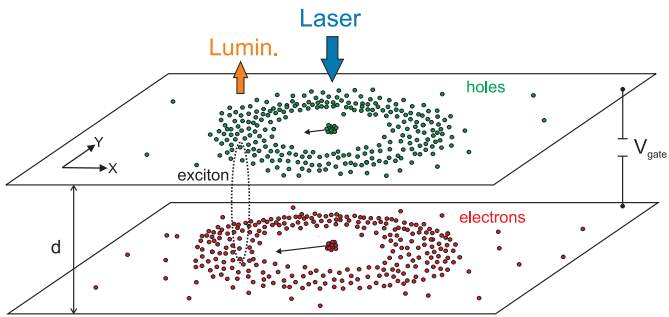}
\caption{\label{fig:fig4} Qualitative schematic of optically-excited
electron-hole bilayer. Both stationary laser pumping in the center
and spatially-distributed luminescence are perpendicular to the
layers. The charge separation between the layers postulated in the
numerical model is due to the external gate voltage $V_{gate}$.}
\end{center}
\end{figure}

To describe the spatial dynamics of $N$ hot electrons and holes, we
use classical equations of motion
\begin{align}
m_{e}^{\ast}\ddot{\mathbf{r}}_{e}^{i}+\gamma_{e}\dot{\mathbf{r}}_{e}^{i} &
={{\displaystyle\sum\limits_{j\neq i}}}\frac{e^{2}(\mathbf{r}_{e}%
^{i}-\mathbf{r}_{e}^{j})}{\left\vert \mathbf{r}_{e}^{i}-\mathbf{r}_{e}%
^{j}\right\vert ^{3}}-{{\displaystyle\sum\limits_{k}}}\frac{e^{2}\left(
\mathbf{r}_{e}^{i}-\mathbf{r}_{h}^{k}\right)  }{\left[  \left(  \mathbf{r}%
_{e}^{i}-\mathbf{r}_{h}^{k}\right)  ^{2}+d^{2}\right]  ^{3/2}},\label{eq1}\\
m_{h}^{\ast}\ddot{\mathbf{r}}_{h}^{i}+\gamma_{h}\dot{\mathbf{r}}_{h}^{i} &
={{\displaystyle\sum\limits_{j\neq i}}}\frac{e^{2}(\mathbf{r}_{h}%
^{i}-\mathbf{r}_{h}^{j})}{\left\vert \mathbf{r}_{h}^{i}-\mathbf{r}_{h}%
^{j}\right\vert ^{3}}-{{\displaystyle\sum\limits_{k}}}\frac{e^{2}%
(\mathbf{r}_{h}^{i}-\mathbf{r}_{e}^{k})}{\left[  (\mathbf{r}_{h}%
^{i}-\mathbf{r}_{e}^{k})^{2}+d^{2}\right]  ^{3/2}},\label{eq2}%
\end{align}
combined with the conditions of exciton formation and optical phonon
emission (see below). Here vectors $\mathbf{r}_{e}^{i}$ and
$\mathbf{r}_{h}^{j}$ are in-plane positions of $i$-th electron and
$j$-th hole ($1\leq i,j\leq N$), $m_{e(h)}^{\ast}$ is the electron
(hole) effective mass, $e$ is the electron charge, and $d$ is the
interlayer distance (see Fig. \ref{fig:fig4}).

In addition to the inertia terms, the left-hand side of
Eqs.(\ref{eq1}-\ref{eq2}) contains phenomenological momentum damping
terms due to the interaction with acoustic phonons with constants
$\gamma_{e(h)}=e/\mu_{e(h)}$, where $\mu_{e(h)}$ is electron (hole)
mobility. The dimensionless equations read
\begin{align}
\ddot{\mathbf{r}}_{e}^{i}+\dot{\mathbf{r}}_{e}^{i} &  ={{\sum\limits_{j\neq
i}}}\frac{(\mathbf{r}_{e}^{i}-\mathbf{r}_{e}^{j})}{\left\vert \mathbf{r}%
_{e}^{i}-\mathbf{r}_{e}^{j}\right\vert ^{3}}-{{\displaystyle\sum\limits_{k}}%
}\frac{\left(  \mathbf{r}_{e}^{i}-\mathbf{r}_{h}^{k}\right)  }{\left[  \left(
\mathbf{r}_{e}^{i}-\mathbf{r}_{h}^{k}\right)  ^{2}+d^{2}\right]  ^{3/2}%
},\label{eq3}\\
\ddot{\mathbf{r}}_{h}^{i}+c_{1}\dot{\mathbf{r}}_{h}^{i} &
={{\displaystyle\sum\limits_{j\neq i}}}\frac{c_{2}(\mathbf{r}_{h}%
^{i}-\mathbf{r}_{h}^{j})}{\left\vert \mathbf{r}_{h}^{i}-\mathbf{r}_{h}%
^{j}\right\vert ^{3}}-{{\displaystyle\sum\limits_{k}}}\frac{c_{2}%
(\mathbf{r}_{h}^{i}-\mathbf{r}_{e}^{k})}{\left[  (\mathbf{r}_{h}%
^{i}-\mathbf{r}_{e}^{k})^{2}+d^{2}\right]  ^{3/2}},\label{eq4}%
\end{align}
with constants $c_{1}=m_{e}^{\ast}\mu_{e}/\left(
m_{h}^{\ast}\mu_{h}\right) $ and $c_{2}=m_{e}^{\ast}/m_{h}^{\ast}$.
Hereafter, we normalized time by
$t_{e}=\sqrt{\epsilon}m_{e}^{\ast}\mu_{e}/e$, where $\epsilon$ is
dielectric constant of the layers, and all distances by $\xi_{e}%
=\sqrt[3]{m_{e}^{\ast}\mu_{e}^{2}}$. To estimate the parameters, we
used well-known experimental values for high-quality undoped
GaAs/AlGaAs QWs. In particular, taking typical $\mu_{e}\sim10^{7}$
cm$^{2}$/(Vs) \cite{mob01,mob02} for temperatures $T\lesssim1$ K,
$m_{e}^{\ast}\approx0.067m_{e}$, $m_{h}^{\ast}\approx0.5m_{e}$
($m_{e}$ is bare electron mass), $\epsilon=12.8$ and
$\mu_{h}\sim0.1\mu_{e}$ \cite{mob}, one gets $t_{e}\sim10^{-9}$ s,
$\xi_{e}\sim10^{-4}$ cm, $c_{1}\sim1$ and $c_{2}\sim0.1$.

The optical phonon emission was modeled in the following way: if the
kinetic energy of a carrier exceeded the energy of optical phonon,
the latter was subtracted from the first and the new direction of
the carrier velocity became random.

\textbf{Simulation of the Laser Pumping.} Stationary optical pumping
of carriers was simulated by generating them in random positions
inside the excitation spot of radius $r_{0}$ with some generation
rate which was modeled in two different ways. In fact, during a MD
simulation the time is changed by discrete steps, with the
elementary time step $\Delta t$. According to the first way
\cite{PRB10, PRB10Er}, the generation rate $p$ was defined as a
probability per $\Delta t$ to create \textit{one} e-h pair in the
excitation spot, so that $p\Delta t<1$. We refer this case as
\textit{single generation regime (SGR)}. Alternatively, one can
consider the formation of several e-h pairs during $\Delta t$. Then
the carrier generation rate (CGR) is defined as a number of e-h
pairs generated in the excitation spot during the time step $\Delta
t$. We call it as \textit{multiple generation regime (MGR)}. Note
that the results of MD simulations differ essentially in the single
and multiple regimes. Indeed, it is intuitively clear that SGR
likely corresponds to weak pumping whereas MGR describes high-power
excitation.

The initial velocities of carriers in the excitation spot were also
chosen randomly within intervals $|\dot{\mathbf{r}}_{e}^{i}|\leq
v_{0}$ and $|\dot{\mathbf{r}}_{h}^{i}|\leq\eta v_{0}$, where we took
$\eta = 0.5$ for all simulations.

During the spatial dynamics of the carriers, the exciton formation
happened if an electron and a hole were close enough to each other,
$\left\vert \mathbf{r}_{e}-\mathbf{r}_{h}\right\vert <a$, where
$a(d)$ is a phenomenological in-layer exciton radius, and their
relative velocity was smaller than some critical value, $\left\vert
\dot{\mathbf{r}}_{e} -\dot{\mathbf{r}}_{h}\right\vert <V_{c}$
\cite{PRB10} (see also Appendix C). Note that the dependence of
exciton formation rate on the e-h relative velocity is one of the
most crucial ingredient for the ring-shaped pattern formation:
assuming the absence of such dependence one always gets a
spatially-monotonic decrease of the luminescence outside the
excitation spot \cite{PRB10}.

To simplify the simulations, we did not consider the exciton
dynamics. It means that as soon as an electron and a hole had formed
an exciton, their dynamics was no longer considered and the position
of the formation event was recorded as a position of photon
emission. Qualitatively, this corresponds to zero exciton lifetime.

One should note that since both the internal and external ring radii
are temperature-independent \cite{But_ring, But_rev}, it is not
advisable to include temperature (i.e., to add a stochastic force in
Eqs.(\ref{eq1}-\ref{eq2})) in the consideration. In turn, since the
low-temperature fragmentation of the external ring \cite{But_ring}
apparently depends on the exciton dynamics, we do not expect to
observe the fragmentation in the simulation results.

Finally, due to the inevitable restrictions in computational power
it was only possible to simulate the dynamics of $N\lesssim10^{4}$
interacting particles. For this reason we had to modify the values
of $a$, $v_{0}$, $V_{c}$, $d$ etc in comparison with realistic
values to facilitate the exciton formation. However, it was clearly
seen that the closer values of those model and real parameters, the
better the correspondence between the MGR simulation results and the
experimental ones. Note that the in-plane motion of the carriers was
not restricted by any spatial boundaries.

\section{Molecular dynamics simulations: Results}

Some preliminary results for the SGR were published in
Ref.\cite{PRB10}. In particular, quasi-1D simulations and the
crucial dependence of the ring pattern formation on the critical
relative velocity $V_{c}$ in the exciton formation condition were
discussed there. In what follows, some new essential results are
described.

\subsection{Single generation regime (SGR)}

The results of MD simulations of Eqs.(\ref{eq3}-\ref{eq4}) in the
SGR for several sets of parameters indicate that there are two
qualitatively different pictures. In general, the in-layer
distribution $n_{\mathrm{lum}}(r)$ of stationary luminescence
exhibits a ring-shaped pattern around the excitation spot. However,
the pattern always contains only one ring. More importantly, the
ring can originate by two qualitatively different ways.

\begin{figure}[!th]
\begin{center}
\includegraphics[width=8.8cm]{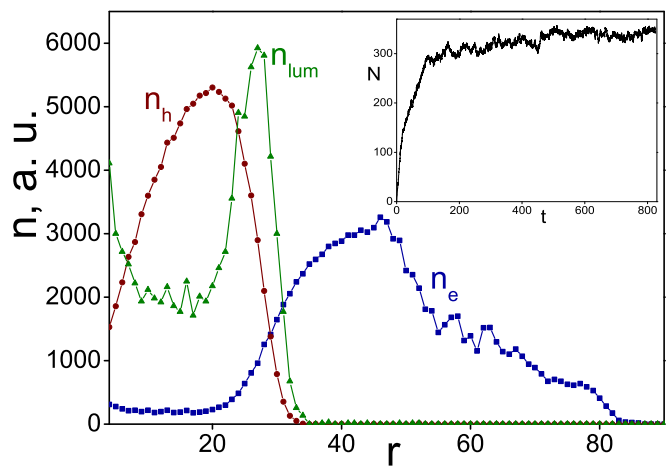}
\caption{\label{fig:fig5} Stationary in-layer distributions of
electrons ($n_{e}$), holes ($n_{h}$) and luminescence
($n_{\mathrm{lum}}(r)$) averaged over time interval (200,830).
Inset: dependence of the total number $N$ of electrons and holes on
time $t$. In a stationary state the carrier generation rate is
balanced by the luminescence rate so $N(t)$ exhibits saturation.
Parameters of the simulation: excitation spot radius $r_{0} = 4$, $p
= 10$, velocity of optical-phonon emission threshold $v_{opt} = 50$,
maximal electron initial velocity $v_{0} = 50$, critical relative
velocity $V_{c} = 10$, critical relative distance $a = 0.2$, first
and second coefficients in the equations of motion for holes $c_{1}
= 1$ and $c_{2} = 0.25$, $d^{2} = 0.01$, time step $\Delta t =
0.0005$.}
\end{center}
\end{figure}

According to the first way, the in-layer distributions of electrons
and holes are separated from each other and the ring occurs in the
region of their overlapping (Fig. \ref{fig:fig5}).

\begin{figure}[!th]
\begin{center}
\includegraphics[width=8.8cm]{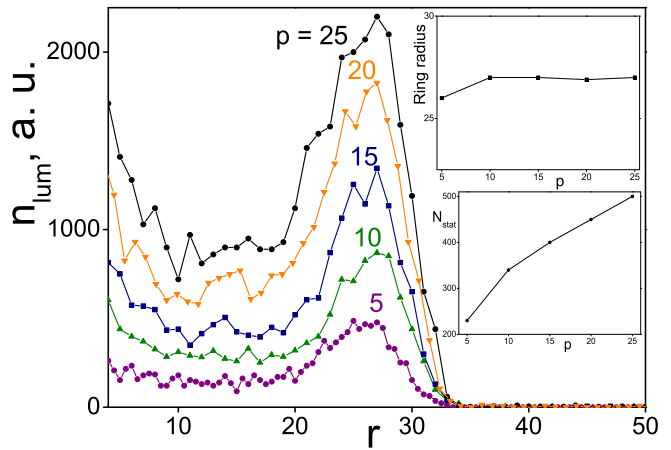}
\caption{\label{fig:fig6} Dependence of stationary in-layer
luminescence distribution $n_{\mathrm{lum}}(r)$ on generation rate
$p$. Top inset: dependence of the luminescence ring radius on $p$.
Bottom inset: stationary total number $N_{stat}$ of electrons and
holes vs $p$. Parameters of the simulations are the same as in Fig.
\ref{fig:fig5}.}
\end{center}
\end{figure}

The dependence of the ring position on generation rate $p$ (Fig.
\ref{fig:fig6}), which mimics the excitation power, shows that
though the luminescence ring intensity increases with $p$ its
position is virtually independent on $p$ (Top inset in Fig.
\ref{fig:fig6}). This behaviour differs from that observed
experimentally (Fig. \ref{fig:fig2}), where the radius of external
luminescence ring grows near linearly with the increase of
photoexcitation intensity and the growth of the internal ring radius
is also quite noticeable. Interestingly, the same behaviour, i.e.,
independence of the ring position on $p$, was observed in quasi-1D
case \cite{PRB10}.

However, there exists an another way of the ring pattern formation.
It was observable at other sets of parameters, in particular, when
maximal initial velocities of the carriers, the critical relative
velocity $V_{c}$ and the distance $a$ were relatively small. Note
that the CGR $p$ was taken in the same range of values as
previously.

\begin{figure}[!th]
\begin{center}
\includegraphics[width=8.8cm]{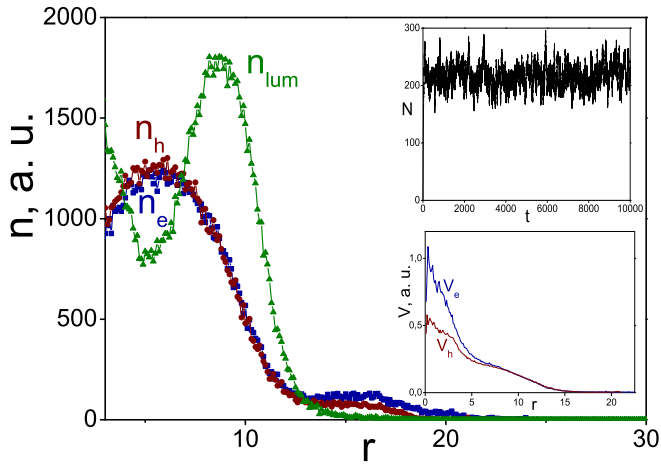}
\caption{\label{fig:fig7} Stationary in-layer distributions of
electrons ($n_{e}$), holes ($n_{h}$) and luminescence
($n_{\mathrm{lum}}(r)$) averaged over time interval (500,10000). Top
inset: dependence of the total number $N$ of electrons and holes on
time $t$ exhibits a stationary state. Bottom inset: dependence of
electron ($V_{e}$) and hole ($V_{h}$) stream velocities vs distance
$r$ from the excitation spot center. Parameters of the simulation:
$r_{0} = 3$, $p = 10$, $v_{opt} = 10$, $v_{0} = 1$, $V_{c} = 1$, $a
= 0.05$, $c_{1} = 1.1$, $c_{2} = 0.3$, $d^{2} = 0.9$ (at smaller
$d^{2}$ the results are essentially the same), time step $\Delta t =
0.005$.}
\end{center}
\end{figure}

\begin{figure}[!th]
\begin{center}
\includegraphics[width=8.8cm]{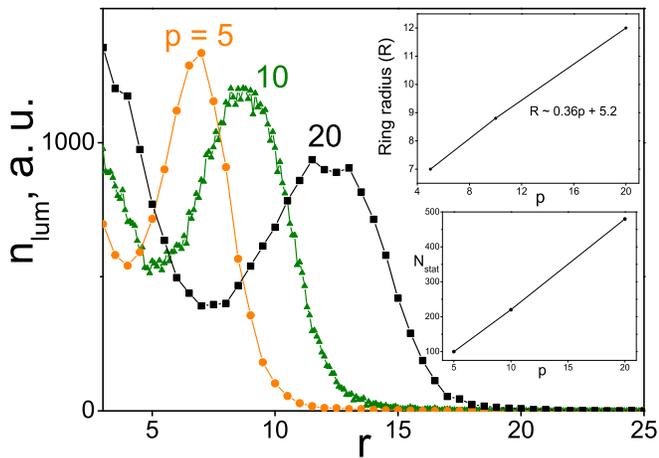}
\caption{\label{fig:fig8} Dependence of stationary in-layer
luminescence distribution $n_{\mathrm{lum}}(r)$ on generation rate
$p$. Top inset: dependence of the luminescence ring radius on $p$.
Bottom inset: stationary total number $N_{stat}$ of electrons and
holes vs $p$. Parameters of the simulations are the same as in Fig.
\ref{fig:fig7}.}
\end{center}
\end{figure}

According the second way, the in-layer distributions of electrons
and holes practically coincide with each other (Fig.
\ref{fig:fig7}). The ring occurs at the outer side of the
distributions, where the differences in the carrier densities and
velocities are small enough to allow the exciton formation. The
dependence of the ring radius on the CGR $p$ (Fig. \ref{fig:fig8})
has shown that, in contrast to the previous case (Figs.
\ref{fig:fig5} and \ref{fig:fig6}), the radius increases linearly
with $p$.

Summarizing the simulation results for the SGR, one can conclude
that the first way of the ring pattern formation, when electron and
hole in-plane distributions are separated and the ring is formed at
their overlapping, does not correspond qualitatively to the
experimental results \cite{But_ring, BLS, But_new}. The second way
could mimic the experimental situation when the excitation power was
such that only the internal ring of luminescence was observable.
However, the position of the ring in the simulations depends on the
pumping rate stronger (linearly) than that of the internal ring (cp.
Fig. \ref{fig:fig2} and Fig. \ref{fig:fig8}) in the experiments
\cite{But_ring, BLS, But_new}. In turn, nearly-linear dependence of
the ring radius on the excitation power is typical for the external
ring of luminescence (Fig. \ref{fig:fig2}), but then the simulations
omit the internal ring.

This misfit can indicate that though the parameters of the
simulations in the second case are closer to realistic ones, one
needs more realistic carrier generation algorithm which would enable
us to model high excitation powers and, at the same time, is
independent on the MD time step $\Delta t$.

\subsection{Multiple generation regime (MGR)}

In the MGR the simulations result in much better correspondence with
the experimental plots (Fig. \ref{fig:fig2}) and, simultaneously,
one can trace some correspondence with the SGR results ("the second
way").

\begin{figure}[!th]
\begin{center}
\includegraphics[width=8.8cm]{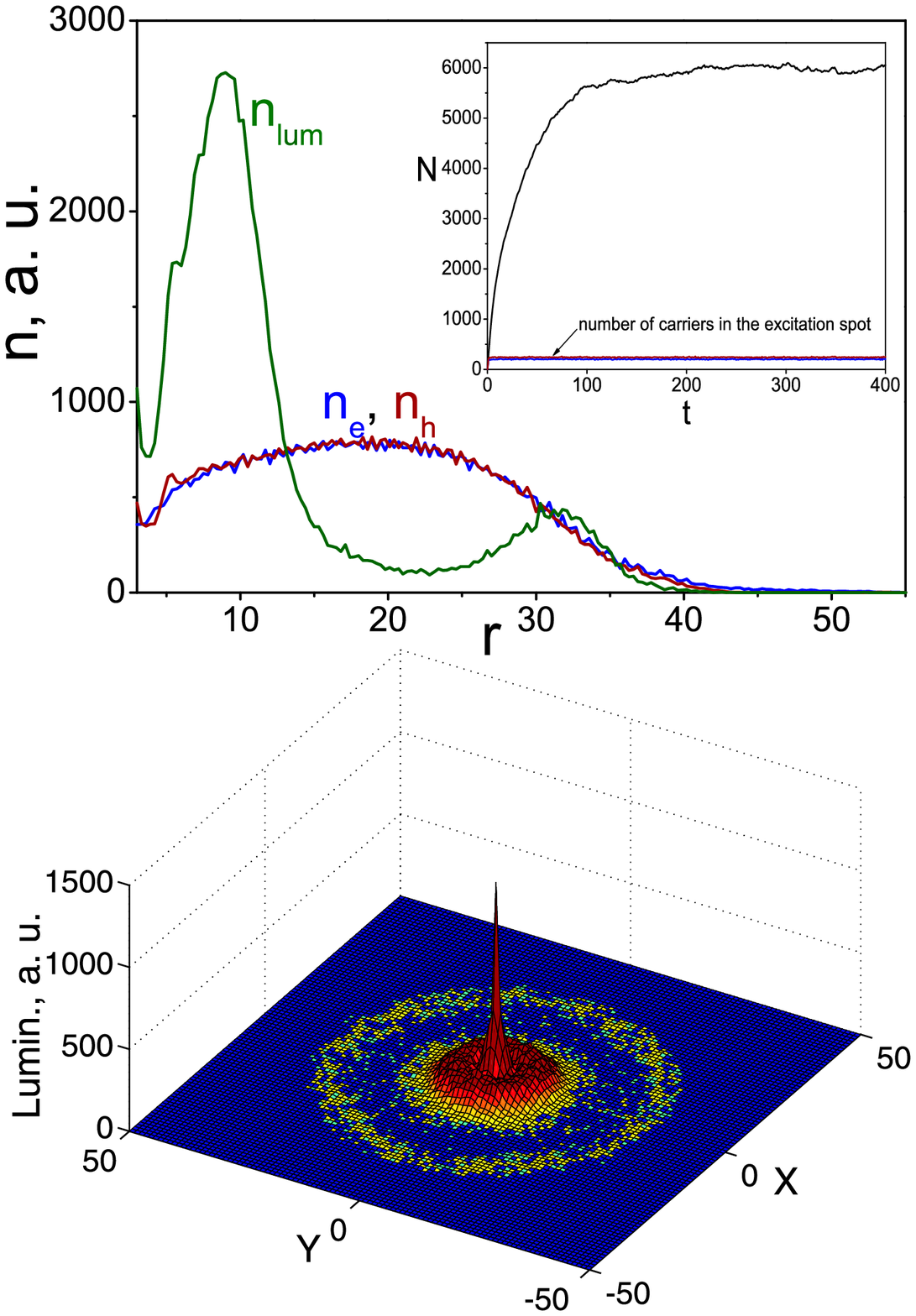}
\caption{\label{fig:fig9} Top: Stationary in-layer distributions of
electrons ($n_{e}$), holes ($n_{h}$) and luminescence
($n_{\mathrm{lum}}(r)$) averaged over time interval (100,400). Inset
on top: dependence of the total number $N$ of electrons and holes on
time $t$ exhibits stationary regime. Bottom: 3D luminescence
pattern. Parameters of the simulation: $r_{0} = 3$, pumping rate
(number of particles generated during time step $\Delta t$ in the
spot) $\Delta N_{spot} = 3$, $v_{opt} = 10$, $v_{0} = 1$, $V_{c} =
1$, $a = 0.05$, $c_{1} = 1.1$, $c_{2} = 0.3$, $d^{2} = 0.9$, $\Delta
t = 0.01$.}
\end{center}
\end{figure}

First, the MGR results show two concentric rings in the in-layer
luminescence pattern: the internal ring has small radius and high
intensity and the external ring has relatively large radius and is
weaker in intensity (Fig. \ref{fig:fig9}). The in-layer
distributions of electrons and holes are similar to those of the
above SGR "second way" (Fig. \ref{fig:fig7}), i.e., the
distributions practically coincide with each other.

Due to the increase of effective pumping power in the MGR case the
total number of carriers in the stationary state (inset in Fig.
\ref{fig:fig9}) is more than one order of magnitude larger than that
in the SGR case. Note that the stationary number of carriers in the
excitation spot is very close to the product of instant CGR ($\Delta
N_{spot}/\Delta t = 300$) and the unit time (= 1). It means that most of
the carriers escape from the spot very fast. This can be understood
by making an estimate of the critical number of carriers in the spot
required for dominant in-layer repulsion (see Section 2): taking
$n_{c}\sim d^{-2}$ one gets $N_{c}= n_{c}\pi r_{0}^{2}\sim\pi\left(
r_{0}/d\right)^{2}\approx30$, which is much smaller than the above
number of carriers formed in the spot during the unit time. (For the
estimate we have used the parameters in Fig. \ref{fig:fig9}
caption.) Thus, the system of charge carriers is in the critical
state.

\begin{figure}[!th]
\begin{center}
\includegraphics[width=8.8cm]{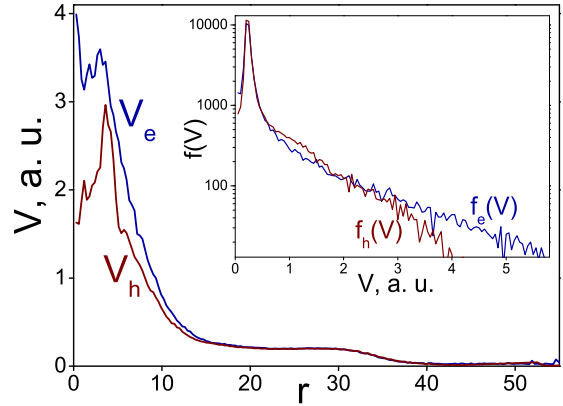}
\caption{\label{fig:fig10} Dependence of electron ($V_{e}$) and hole
($V_{h}$) stream velocities vs distance $r$ from the excitation spot
center. One can see that the carrier velocities within the
excitation spot region essentially exceed the maximal initial
velocity (= 1). Note that $V_{e}$ and $V_{h}$ are the velocities of
that part of the carriers which are below threshold of optical
phonon emission. Inset: Electron ($f_{e}(V)$) and hole ($f_{h}(V)$)
distributions over single-particle velocities. Parameters of the
simulation are the same as in Fig. \ref{fig:fig9}.}
\end{center}
\end{figure}

The mechanism of the ring pattern formation in these conditions is
as follows. The internal ring is formed due to the carriers which
have emitted optical phonons. It can be seen in Fig. \ref{fig:fig10}
by the sharp changes of carrier velocities within the excitation
spot. Note that the velocities essentially exceed the maximal
initial velocities there. (In addition, the optical phonon emissions
were seen during the simulations by changes of instant maximal
velocities of the carriers that were up to the threshold velocity of
optical phonon emission.) It means that the in-layer repulsive
Coulomb interaction is dominant in the excitation spot. The
repulsive forces accelerate the carriers in such a way that the major part of
them emits optical phonons and then quickly forms excitons. The
carriers with velocities beneath the threshold velocity of optical
phonon emission go further emitting acoustic phonons and,
eventually, form excitons relatively far away from the excitation
spot. This description is in excellent agreement with all sets of
simulations in the MGR so it justifies the second scenario suggested
in Section 2. Some details of the Coulomb repulsion accompanied by
phonon emission are given in Appendix B.

\begin{figure}[ptbh]
\begin{center}
\includegraphics[width=8.8cm]{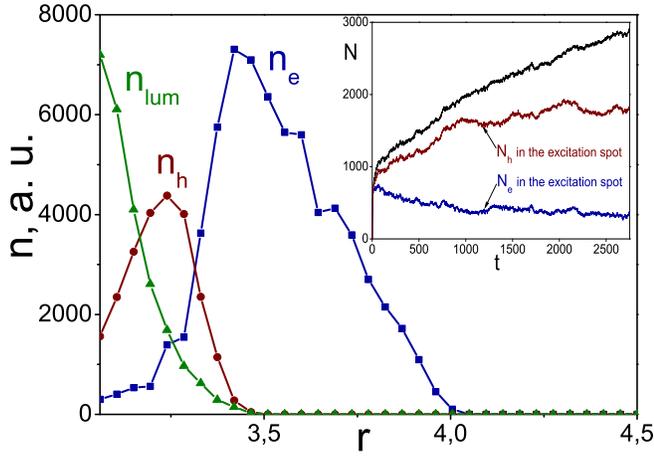}
\caption{\label{fig:fig11} In-layer distributions of electrons
($n_{e}$), holes ($n_{h}$) and luminescence ($n_{\mathrm{lum}}(r)$)
averaged over time interval (1500,2700) in the absence of any direct
Coulomb interactions. Inset: dependence of the total number $N$ of
electrons and holes on time $t$. It shows that the dynamics is still
non-stationary. Parameters of the simulation are the same as in Fig.
\ref{fig:fig9}.}
\end{center}
\end{figure}

To prove the crucial role of Coulomb interactions, we have performed
the simulation where the interactions are switched off even though
the exciton formation condition holds. The results are shown in Fig.
\ref{fig:fig11}. One can see that the saturation of $N(t)$ (i.e.,
the stationary state) is absent during the time which essentially
exceeds previous simulation times for the MGR. The in-layer
distributions of electrons and holes resemble those for the "first
way" SGR, i.e., the distributions are spatially separated.
However, the ring does not form and the luminescence decreases
monotonically from the center.

We now discuss the dependence of the ring-shaped pattern on the
pumping power and the critical relative velocity $V_{c}$ from the
exciton formation condition.

\begin{figure}[ptbh]
\begin{center}
\includegraphics[width=8.8cm]{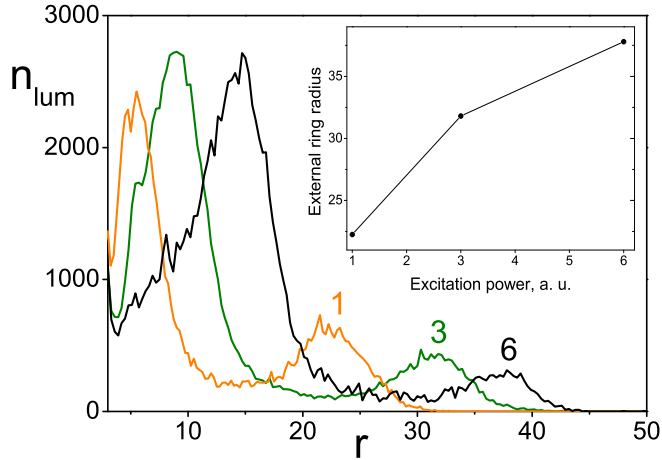}
\caption{\label{fig:fig12} Dependence of stationary in-layer
luminescence distribution $n_{\mathrm{lum}}(r)$ on excitation power
with the number $\Delta N_{spot}$ of particles generated per time
step in the spot shown by numbers (1, 3, 6) near the corresponding
curves. Inset: dependence of the external ring radius on excitation
power. Other parameters of the simulations are the same as in Fig.
\ref{fig:fig9}.}
\end{center}
\end{figure}

The dependence of in-plane positions of the rings on the pumping
rate is shown in Fig. \ref{fig:fig12}. The external ring radius
increases nearly linearly whereas the internal ring radius grows
more slowly. This behaviour exhibits a good agreement with the
experimental curves (Fig. \ref{fig:fig2}). (To determine the
dependences more accurately one needs to collect larger statistics
that is a very time-consuming procedure due to the large total
numbers ($\sim 10^{4}$) of particles in the MGR.)

\begin{figure}[ptbh]
\begin{center}
\includegraphics[width=8.8cm]{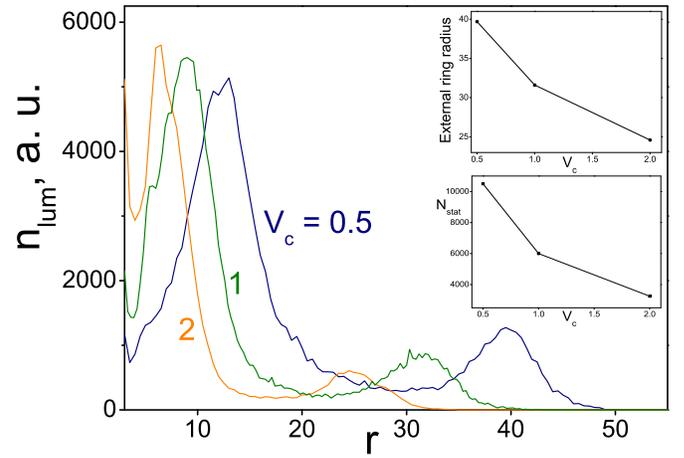}
\caption{\label{fig:fig13} Dependence of stationary in-layer
luminescence distribution $n_{\mathrm{lum}}(r)$ on critical relative
velocity $V_{c}$ from the exciton formation condition. Top inset:
dependence of the external ring radius on $V_{c}$. Bottom inset:
stationary total number $N_{stat}$ of carriers vs $V_{c}$. The
pumping rate $\Delta N_{spot} = 3$. Other parameters of the
simulations are the same as in Fig. \ref{fig:fig9}.}
\end{center}
\end{figure}

The dependence of the ring-shaped luminescence pattern on the
critical relative velocity $V_{c}$ shows (Fig. \ref{fig:fig13}) that
the smaller $V_{c}$ the larger the external ring radius (top inset
in Fig. \ref{fig:fig13}) and the ring intensity. At the same time
the stationary number of carriers grows up to the limiting values of
order of $10^{4}$ (bottom inset in Fig. \ref{fig:fig13}) for the
available computational power.

\begin{figure}[ptbh]
\begin{center}
\includegraphics[width=8.8cm]{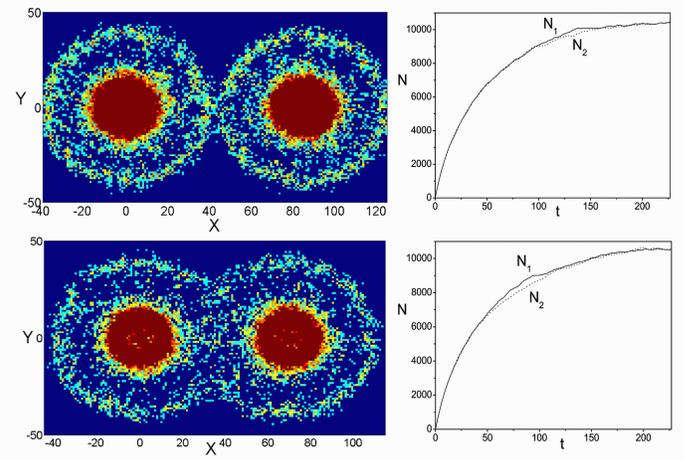}
\caption{\label{fig:fig14} Left: Luminescence patterns for the case
of two excitation spots at two distances, 85 (top) and 70
(bottom), between the spots. Right: The corresponding dependences of
total numbers $N_{1,2}$ of carriers excited by the spots on time $t$
show that the system arrives at the stationary state. Parameters of
the simulations are the same as in Fig. \ref{fig:fig9} except
critical relative velocity, here $V_{c} = 0.5$.}
\end{center}
\end{figure}

Finally, we have performed simulations with two identical but
spatially separated excitation spots to compare our results (see
Fig. \ref{fig:fig14}) with experimental pictures \cite{cond-mat,
BLS}. One can see that when the spots are placed close enough, the
external rings open towards each other forming a figure similar to
one of the Cassini ovals. This behaviour also corresponds to the
experiments even though the simulations do not include the exciton
dynamics.

Summarizing the simulation results for the MGR, we conclude that
they show reasonable correspondence with the experiments
\cite{But_ring, cond-mat, BLS, But_new}. Therefore the theoretical
explanations \cite{BLS,Rap04,Den,Ha,Sug06,But_new} of the ring
pattern formation based on the diffusion-induced spatial separation
of in-layer distributions of electrons and holes (see Appendix A)
must be revised.

\section{Conclusion and discussion}

It has been shown that the stationary ring-shaped luminescence
pattern forms due to the hot carrier transport (HCT) caused by the
in-plane electric fields which appear at high enough excitation
power in the excitation spot region. The HCT is essentially
non-diffusive. In particular, the internal luminescence ring appears
due to electrons and holes emitting optical phonons, whereas the
external ring forms due to the relaxation of the carriers that are
initially below the threshold of optical phonon emission. To form
excitons, these carriers relax emitting acoustic phonons and, in
addition, due to the interlayer Coulomb drag.

The ring-shaped pattern formation is particularly interesting as a
possible bright signature of self-organized criticality \cite{Bak,
Bak-book}. Though "the second scenario" naturally involves the SOC
regime, the MGR simulations reported have been performed in the
critical state. So the transition to the SOC regime as well as its
statistical properties (e.g., $1/f$-noise) in this system are still open questions.

The author thanks L. P. Paraskevova and L. I. Kondrashova for the
encouragement, and Yu. M. Kagan and F. V. Kusmartsev for helpful discussions.

\section{Appendixes}

\subsection{Appendix A: Diffusive model of charge carrier transport}

The diffusive transport model \cite{BLS} used to explain the
experiments \cite{But_ring} was based on two reaction-diffusion
equations
\begin{align}
\dot{n}_{e} &  =D_{e}\nabla^{2}n_{e}-wn_{e}n_{h}+J_{e}\left(  r\right)  ,\label{BL1}\\
\dot{n}_{h} &  =D_{h}\nabla^{2}n_{h}-wn_{e}n_{h}+J_{h}\left(
r\right)  ,\label{BL2}
\end{align}
where $n_{e}$ and $n_{h}$ are electron and hole two-dimensional (2D)
densities, $w$ is electron-hole binding rate to form an exciton. The
source term $J_{h}\left(
r\right)=P_{ex}\delta\left(\mathbf{r}\right)$ for photoexcited holes
is focused in the local excitation spot. The density of photoexcited
electrons is supposed to be negligible in comparison with
equilibrium electron density $n_{\infty}$ in the absence of laser
excitation. When $n_{\infty}$ is spatially disturbed due to the
presence of holes, there appears the electron current $J_{e}\left(
r\right)  =I-a\cdot n_{e}\left( r\right)$, which is spread in the quantum-well
plane. Here, $I$ and $a\cdot n_{e}$ are the currents in and out of
the system, respectively, such that $n_{\infty }=I/a$. (Note that $a$ here is not
the critical relative distance used in the simulations but an
independent parameter.) Implying the stationary regime and the
symmetry over the polar angle, and neglecting the exciton diffusion
\cite{Levitov}, one gets the exciton PL intensity $I_{PL}(r)\propto
n_{e}(r)n_{h}(r)$.

\begin{figure}[!t]
\begin{center}
\includegraphics[width=8.8cm]{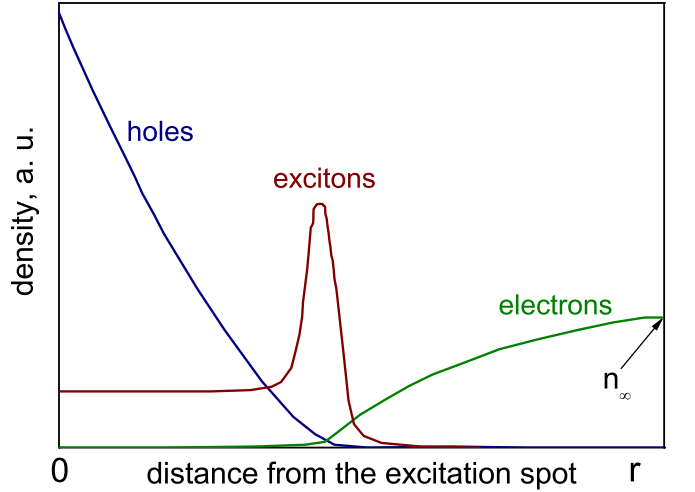}
\caption{\label{fig:fig15} The density distributions (the scales are
in a.u.) obtained in Ref.\protect\cite{BLS} with the use of the
diffusive transport model described in text.}
\end{center}
\end{figure}

Authors of Ref.\cite{BLS} have assumed that a luminescence ring with
radius $R$ appears at the overlap of the electron and hole densities
(see Fig. \ref{fig:fig15}) so that $n_{h}>>n_{e}$ at $r<R$ and
$n_{h}<<n_{e}$ at $r>R$ with $n_{e}\left( r\rightarrow\infty\right)
= n_{\infty }$ (Fig. \ref{fig:fig15}).
Neglecting the exciton formation term $wn_{e}n_{h}$ far from the
boundary $r=R$, one gets for holes
\begin{equation}
\nabla^{2}n_{h}\equiv\frac{d^{2}n_{h}}{dr^{2}}+\frac{1}{r}\frac{dn_{h}}%
{dr}=-\left(  P_{ex}/D_{h}\right)  \delta\left(\mathbf{r}\right).
\end{equation}
Using boundary condition $n_{h}\left(r=R\right)=0$ it results in
\begin{equation}
n_{h}\left(  r\leq R\right) =\frac{P_{ex}}{2\pi D_{h}}\ln\left(  \frac{R}%
{r}\right)  .
\end{equation}
For electrons one gets
\[
\nabla^{2}n_{e}=-\left(  I/D_{e}\right)  +\left(  a/D_{e}\right)
\cdot n_{e}\left(  r\right)
\]
with boundary conditions $n_{e}\left(  r=R\right)  =0,$ \
$n_{e}\left( r\rightarrow\infty\right)  =n_{\infty}.$

Denoting $\Delta=\left(  n_{e}-n_{\infty}\right)  /n_{\infty}$ and\
$x=r/\lambda$, where
$\lambda\equiv\sqrt{D_{e}/a}=\sqrt{D_{e}n_{\infty
}/I}$ is a characteristic length, we arrive at%
\begin{equation}
\frac{d^{2}\Delta}{dx^{2}}+\frac{1}{x}\frac{d\Delta}{dx}-\Delta=0.
\end{equation}
The last equation is the modified Bessel equation of zero order. The
solution is
\begin{equation}
\Delta\left(  x\right)  =A\cdot K_{0}\left(  x\right)  ,
\end{equation}
where $A$ is some constant and $K_{0}\left(  x\right)  $ is modified
Hankel (or MacDonald) function of zero order, such that $K_{0}\left(
x\rightarrow 0\right)  \approx\ln\frac{1}{x},$ \ $K_{0}\left(
x\rightarrow\infty\right) =0$. Thus,
\begin{equation}
n_{e}\left(  r\right)  /n_{\infty}=1+A\cdot K_{0}\left(
r/\lambda\right)  .
\end{equation}
If $\lambda\gg R$, as supposed in \cite{BLS}, then at $R<r<<\lambda$
the electron density $n_{e}\left(  r\right)
/n_{\infty}\approx1+A\ln\left( \lambda/r\right)$. Using the boundary
condition $n_{e}\left(  r=R\right)=0$ one finds coefficient $A$.
Finally,
\begin{equation}
n_{e}\left(  R<r<<\lambda\right)  \approx n_{\infty}\left[
1-\ln\left( \lambda/r\right)  /\ln\left(  \lambda/R\right)  \right]
.
\end{equation}
Then it has been assumed that at the boundary between the electron
and hole densities the total current is zero, i.e.,
\begin{equation}
D_{e}\frac{\partial n_{e}}{\partial r}|_{r=R}=-D_{h}\frac{\partial n_{h}%
}{\partial r}|_{r=R}.\label{Boundary}
\end{equation}
From here it follows that
\begin{equation}
D_{e}n_{\infty}/\ln\left(  \lambda/R\right)  =P_{ex}/2\pi,
\end{equation}
so the ring radius can be expressed as
\begin{equation}
R=\lambda\exp\left(  -2\pi D_{e}n_{\infty}/P_{ex}\right)  .\label{rad}%
\end{equation}
If we set $D_{h}=0$, then the ring radius $R$ must be equal to zero
since, according to the model \cite{BLS}, the diffusion of holes is the only
reason why they move out of the excitation spot. However,
Eq.(\ref{rad}) does not depend on $D_{h}$ so the ring radius is not
zero at $D_{h}=0$, i.e., the ring exists even if all holes are left
into the excitation spot. This clearly unphysical result
is not a consequence of the limiting case $\lambda\gg R$, but rather
comes from wrong initial assumptions. (In Ref.\cite{Ha} the result,
Eq.(\ref{rad}), has been generalized, but even then $R$ does not
depend on $D_{h}$.)

We note that in the original paper \cite{BLS} the erroneous formula for
$R$ has been given \cite{com}:
\begin{equation}
R_{orig}=\lambda\exp\left(  -2\pi D_{e}n_{\infty}/\left(
D_{h}P_{ex}\right) \right).
\end{equation}
The presence of $D_{h}$ in the exponent denominator could be
deceiving since at first sight (i.e., without dimensionality check:
$[D_{e}]=[D_{h}]=cm^{2}/s$, $[n_{\infty}]=cm^{-2}$,
$[P_{ex}]=s^{-1}$) it looks reasonable.

In addition, both the DTM \cite{BLS} and its modifications
\cite{Rap04,Den,Ha,Sug06} could not explain in principle why the
external luminescence ring appears only when the excitation power
exceeds some critical value.

Nevertheless, the drift-diffusion regime can be applicable for slow
charge carriers near the luminescence ring at $r\sim R$. (As before,
we do not consider equilibrium carriers and are only focused on
photogenerated ones.) In particular, the continuity equations in
this regime are given by
\begin{align}
\dot{n}_{e(h)}+\operatorname{div}\mathbf{i}_{e(h)}  &  =g_{e(h)}%
-\Gamma,\label{2}\\
\dot{n}_{X}+\operatorname{div}\mathbf{i}_{X}  &
=\Gamma-n_{X}/\tau_{X}.
\label{3}%
\end{align}
Here $n_{e}$, $n_{h}$ and
$\mathbf{i}_{e}=-n_{e}\mu_{e}\mathbf{E}-D_{e}\nabla n_{e}$,
$\mathbf{i}_{h}=n_{h}\mu_{h}\mathbf{E-}D_{h}\nabla n_{h}$ are 2D
densities and particle flux densities of uncoupled electrons in
plane $z=d/2$ and holes in plane $z=-d/2$, $\mu_{e(h)}$ is electron
(hole) mobility. The particle flux density for excitons
$\mathbf{i}_{X}\approx-D_{X}\nabla n_{X}$, where $n_{X}$ is the
interlayer exciton density. The contribution from the dipole-dipole
interaction between the excitons is omitted in $\mathbf{i}_{X}$
since it appears as an above-linear correction on $n_{X}$. The
carrier generation rates $g_{e(h)}\left( \mathbf{r},t\right)$ are
some given functions. The exciton formation rate can be written as
(hereafter inessential constant prefactors are dropped)
\begin{equation}
\Gamma\left(  \mathbf{r},t\right)  =%
{\displaystyle\int}
w\left(  \left\vert \mathbf{v}_{1}-\mathbf{v}_{2}\right\vert \right)
f_{e}\left(  \mathbf{r},\mathbf{v}_{1},t\right)  f_{h}\left(  \mathbf{r}%
,\mathbf{v}_{2},t\right)  d^{2}\mathbf{v}_{1}d^{2}\mathbf{v}_{2}, \label{4}%
\end{equation}
where $f_{_{e(h)}}\left(  \mathbf{r},\mathbf{v},t\right)  $ is the
electron (hole) distribution function, so that $n_{e(h)}\left(
r,t\right)  ={\int }f_{_{e(h)}}\left(  r,\mathbf{v},t\right)
d^{2}\mathbf{v}$, and $w\left( v\right)$ is the specific exciton
formation rate. The exciton lifetime $\tau_{X}$ is supposed to be
density-independent. Finally, the Poisson equation for the electric
field reads (time dependence is dropped; $\epsilon$ is dielectric
constant)
\begin{align}
\operatorname{div}\left(  \epsilon\mathbf{E}(r,z)\right) & =4\pi
e[\left(  n_{h}(r)+n_{X}(r)\right)  \delta(z+d/2)-\label{5}\\
&  -\left(  n_{e}(r)+n_{X}(r)\right)  \delta(z-d/2)].\nonumber
\end{align}
It includes the contribution of the interlayer exciton dipole fields
and keeps the electroneutrality for the free carrier system when the
exciton formation is suppressed ($n_{X}(r)=0$).

At $r\sim R$ one can put $w\left(  v\right)  \approx w_{\max}$, then
$\Gamma\left(  r\right)  \approx w_{\max}n_{e}\left(  r\right)
n_{h}\left( r\right)  $ and the Eqs.(\ref{2}),(\ref{3}),(\ref{5})
with $g_{e(h)}=0$ become a closed system.

We note that the ambipolar electric field $\mathbf{E}$ might play an
important role in the formation of a sharp intensity profile of the
external luminescence ring. In this regard, it is useful to note that
the FWHM of the external ring intensity is almost independent on the
ring radius $R$ at high excitation powers (see Fig. \ref{fig:fig2}).

\subsection{Appendix B}

If $e^{2}\sqrt{n}>\hbar\omega_{opt}$ at $n>d^{-2}$ the in-layer
Coulomb repulsion in the excitation spot might be "exhausted" at
small distances: the potential energy of the carriers transforms
into kinetic one which, in turn, is spent for the fast optical
phonon emission so that the carriers do not go far from the
excitation spot.

To estimate whether the values of carrier densities in the
excitation spot are sufficient for such process, let us consider two
electrons resting at distance $r_{0}$ from each other at moment
$t=0$. We neglect the energy dissipation due to acoustic phonon
emission first, to facilitate the effect described above. Then the
equation of motion is
\[
m\ddot{r}=e^{2}/r^{2},
\]
with $m=m_{e}^{\ast}/2,$ $r=\left\vert \mathbf{r}_{1}-\mathbf{r}%
_{2}\right\vert $. (Dielectric constant $\epsilon$ is introduced by
$e^{2}\rightarrow e^{2}/\epsilon$ in the final expression.) The solution is
expressed through the inverse function,
\[
t/t_{0}=\sqrt{x^{2}-x}+\frac{1}{2}\ln\left(  x+\sqrt{x^{2}-x}\right)  ,
\]
where $x=r/r_{0}$, $t_{0}=\sqrt{mr_{0}^{3}/\left(  2e^{2}\right)}$.
At $t>(3\div4)t_{0}$ with a good precision the velocity $v\approx
v_{\infty }=r_{0}/t_{0}$, i. e., the electrons move near uniformly
at large times. The condition of optical phonon emission reads
\[
mv^{2}/2=e^{2}/r_{0}-e^{2}/r\geq\hbar\omega_{opt},
\]
from where it follows
\[
r\geq r_{c}=\frac{r_{0}}{1-\hbar\omega_{opt}/\left(
e^{2}/r_{0}\right)}.
\]
Substituting $r_{0}\sim n^{-1/2}$ in the expression for $r_{c}$, one
has
\[
n\sim\left(\epsilon\hbar\omega_{opt}/e^{2}+1/r_{c}\right)^{2}
\]
The value $n_{c}^{*}=\left(
\epsilon\hbar\omega_{opt}/e^{2}\right) ^{2}\approx4\cdot10^{13}$
cm$^{-2}$ ($\epsilon\approx12.8,$ \ $\hbar \omega_{opt}\approx37$
meV for GaAs) at $r_{c}=\infty$ is the smallest density at which the
process of optical phonon emission is dominant. Note that at
$n>n_{c}^{*}$ the effect is extremely pronounced, e.g., at
$n=1.003n_{c}^{*}$ (extra 0.3\% to $n_{c}^{*}$) one gets
$r_{c}\approx1$ $\mu$m. At $n=n_{c}^{*}$, we obtain
$r_{0}\sim0.1a_{B}$ for GaAs.

To find a qualitative dependence of carrier flux velocity on
in-plane coordinates when the carrier kinetic energies are below the
threshold of optical phonon emission, we consider the previous
model adding the dissipation due to acoustic phonons. Then for two
electrons we have (unit vectors for the Coulomb force are dropped)
\begin{align*}
m_{e}^{\ast}\mathbf{\ddot{r}}_{1}  &  =e^{2}/\left\vert \mathbf{r}%
_{1}-\mathbf{r}_{2}\right\vert ^{2}-\gamma\mathbf{\dot{r}}_{1},\\
m_{e}^{\ast}\mathbf{\ddot{r}}_{2}  &  =-e^{2}/\left\vert \mathbf{r}%
_{1}-\mathbf{r}_{2}\right\vert ^{2}-\gamma\mathbf{\dot{r}}_{2},
\end{align*}
where $\gamma=e/\mu_{e}$ is dissipation coefficient and $\mu_{e}$ is
electron mobility. Substituting $\mathbf{R}=\left( \mathbf{r}_{1}+\mathbf{r}%
_{2}\right)/2$ and $\mathbf{r}=\mathbf{r}_{1}-\mathbf{r}_{2}$, in
the center-of-mass frame ($\mathbf{\dot{R}}=0$), the equation of
motion reads
\[
m\ddot{r}=e^{2}/r^{2}-\frac{1}{2}\gamma\dot{r}.
\]
The dissipation of energy $E(r)=mv^{2}\left(  r\right)  /2+e^{2}/r$
is given by $dE/dr=-\gamma v\left(r\right)/2$ (since $dE/dt=-\gamma
v^{2}/2$) that leads to equation
\[
mv\left(  r\right)  \frac{dv}{dr}=\frac{e^{2}}{r^{2}}-\frac{1}{2}\gamma
v\left(  r\right)  .
\]
In the dimensionless form ($r=r_{0}x,$ $v=v_{\infty}u$), we have
\begin{equation}
u(x)\left[  \frac{du}{dx}+a\right]  =\frac{1}{2x^{2}}, \label{Abel}%
\end{equation}
where parameter $a=\sqrt{\gamma^{2}r_{0}^{3}/\left( 8me^{2}\right)
}$, $x\geq1$ and $u(x=1)=0$. (To avoid a confusion, note again that
this $a$ has its own meaning.) Using relation $\gamma=e/\mu_{e}$,
one can rewrite the parameter as $a=\frac{1}{2}\left(
r_{0}/\xi_{e}\right) ^{3/2}$, where $\xi
_{e}=\sqrt[3]{m_{e}^{\ast}\mu_{e}^{2}}$ is characteristic length
scale. Taking for GaAs $\xi _{e}\sim1$ $\mu m$ (see the description
of the numerical model in the main text) and $r_{0}\sim
n_{c}^{-1/2}\sim d\sim10^{-6}$ cm, one gets $a\sim10^{-3}$.
\begin{figure}[!th]
\begin{center}
\includegraphics[width=8.75cm]{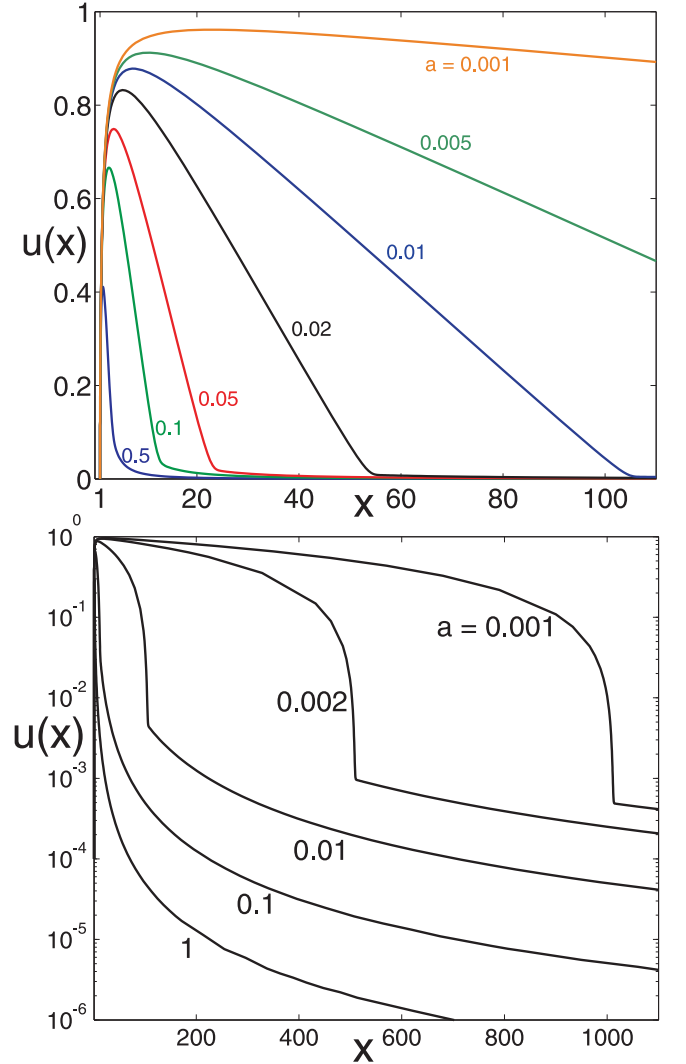}
\caption{\label{fig:fig16} Numerical solutions of Eq. (\ref{Abel})
with $u(x)=v/v_{\infty}$ and $x=r/r_{0}$ for different values of
parameter $a$. Top: Linear scale. Bottom: Log scale.}
\end{center}
\end{figure}
The equation (\ref{Abel}) can be reduced to the Abel equation of the
second kind. Its exact analytical solution is unknown. However, a
qualitative behaviour of $u(x)$ can be found from asymptotic
solutions at $x\rightarrow1$ and $x\rightarrow\infty$ (see also
numerical solutions in Fig. \ref{fig:fig16}). It is natural to
assume that velocity $u(x=\infty)=0$. So $du/dx$ can be dropped in
Eq. (\ref{Abel}) at $x\rightarrow\infty$ that gives
$u_{\infty}(x)\approx 1/(2ax^{2})$. Let us now rewrite Eq.
(\ref{Abel}) as
\[
\frac{du}{dx}=\frac{1}{2x^{2}u}-a.
\]
At $x\rightarrow1$, due to the initial condition $u\rightarrow0$ so now the
last term in the right-hand side can be dropped. This gives $u_{1}%
(x)\approx\sqrt{1-1/x}$ at $x\approx1$. The dependence of $u(x)$ on
$a$ is as follows: the larger $a$ the narrower the width of the peak
and the smaller its height (see Fig. \ref{fig:fig16}). Note that in
the laboratory frame $v_{e,h}=v/2$ and $r_{e,h}=r/2$.

It is also interesting to note that the MGR simulations indicate
that the external luminescence ring appears only if the carrier
stream velocities $V_{e(h)}(r)$ as functions of distance $r$ from
the excitation spot (see Fig. \ref{fig:fig10}) have a non-monotonic
dependence similar to that in Fig. \ref{fig:fig16}, i.e., at small
$r$ the velocities grow, reaching the maximal values at some finite
$r$ (often within the excitation spot region), and then they decrease
to zero at large $r$.

\subsection{Appendix C: Suppression of exciton formation
in the bilayer at high e-h relative velocity}

Here, we illustrate how the critical relative velocity
$V_{c}$ appears in principle in the exciton formation condition. We
can distinguish two mechanisms referred further as "geometric" and
"kinetic" ones, which lead to the existence of the critical e-h
relative velocity above which the interlayer exciton formation is
strongly suppressed.

\textbf{1. "Geometric" mechanism.} Due to the bilayer geometry, at
sufficiently large relative e-h velocity the interlayer Coulomb
attraction, which results in an exciton formation, is suppressed. To
show this, let us consider the Fourie transform $U_{q}=%
{\displaystyle\int}
d^{2}\mathbf{r}\exp\left(  i\mathbf{qr}\right)  U\left(  r\right)  $
of the pair interaction potential $U\left(  r\right)
=-e^{2}/\sqrt{r^{2}+d^{2}}$ between an electron from one layer and a
hole from another ($r$ is in-plane relative distance). One gets
\begin{equation}
U_{q}=-2\pi e^{2}%
{\displaystyle\int\limits_{0}^{\infty}}
\frac{J_{0}\left(  qr\right)  rdr}{\sqrt{r^{2}+d^{2}}}=-\frac{2\pi e^{2}}%
{q}\exp\left(  -qd\right)  , \label{Co}%
\end{equation}
the "screened Coulomb potential" in the momentum space. (The effects
of charge screening studied in the random phase approximation
\cite{screen} lead to a change in the preexponential factor, which
is not important in this consideration.) Thus, if the electron-hole
relative velocity $V=\hbar q/m>V_{c}=\hbar/\left( md\right)$, the
interaction between the carriers decreases exponentially with the
increase of $V$. This means that one can neglect the interaction as
well as the exciton formation at $V>V_{c}$. At $d\approx10^{-6}$ cm
\cite{But_rev} and reduced e-h mass $m\approx0.06m_{e}$ in GaAs, we obtain
$V_{c} \approx3\cdot10^{7}$ cm/s, which is of the same order of
magnitude as the threshold velocity $v_{max}$ of optical phonon emission
in GaAs ($v_{max}/V_{c}\approx1.5$).

\textbf{2. "Kinetic" mechanism.} The second mechanism is based on
the fact that to form an exciton the unbound electron-hole pair must
emit acoustic phonon. (Here we suppose that the carrier velocities
are below the threshold of optical phonon emission.)

To illustrate this, let us consider a model system: an infinite
train of electrons separated from each other by distance $L$
uniformly moves with velocity $V$ along a thread and an immovable
hole is located at distance $d$ from the thread. The interaction
potential between the electron train and the hole as a function of
time reads
\[
U(t)=-{\displaystyle\sum\limits_{k=-\infty}^{+\infty}}\frac{e^{2}}{\rho_{k}},
\]
where $\rho_{k}=\sqrt{\left(kL+Vt\right)^{2}+d^{2}}$. Although the
sum diverges as $1/\left\vert k\right\vert$, the relative value of
the potential $\delta U(t)=U(t)-U(0)$ is convergent. The components
of the corresponding force acting on the hole along and
perpendicular to the thread are given by
\[
F_{\parallel}(t)= {\displaystyle\sum\limits_{k=-\infty}^{+\infty}}%
\frac{e^{2}\left(  kL+Vt\right)}{\rho_{k}^{3}},\text{ \ }F_{\perp}(t)=%
{\displaystyle\sum\limits_{k=-\infty}^{+\infty}}\frac{e^{2}d}{\rho_{k}^{3}}.
\]
Both the potential and the force are periodic functions of time with
period $T=L/V$, e.g., $\delta U(t+T)=\delta U(t)$.

In the 2D case, when there exists a relative flow (with velocity $V$) of
electrons in one layer and holes in another, the interlayer
interaction potential between the electron flow and a given hole
oscillates with frequency $\omega\sim V\sqrt{n}$, where $n$ is 2D
density of electrons in the flow. If this frequency is higher than
$\tau^{-1}$, where $\tau=\min(\tau_{e-ac},\tau_{h-ac})$ is the
minimal carrier-acoustic phonon scattering time, the exciton
formation in real space is not possible. Thus, one can write the
condition of exciton formation as
\begin{equation}
V<V_{c}\sim1/\sqrt{n\tau^{2}}. \label{condit}
\end{equation}
If the carrier densities are essentially different, one should take
$n=\max(n_{e},n_{h})$ in (\ref{condit}). At $n\sim10^{10}$ $cm^{-2}$
and $\tau\sim10^{-9}$ s the critical relative velocity is
$V_{c}\sim10^{4}$ cm/s.

In fact, one can get the estimate (\ref{condit}) in a more simple
way. We suppose for definiteness that locally $n_{e}>n_{h}$.
Then an e-h pair with relative velocity $V$ can form an exciton if
\[
V\tau<\bar{r}_{e-e}\sim n_{e}^{-1/2},
\]
that is the electron or the hole should have time to emit an
acoustic phonon before the next electron would come to the hole. The
condition (\ref{condit}) follows directly from the last formula.


\begin{thebibliography}{37}

\bibitem {Mos} S. A. Moskalenko, Sov. Phys. Solid State \textbf{4}, 199 (1962).

\bibitem {KK}L. V. Keldysh, A. N. Kozlov, Sov. Phys. JETP \textbf{27}, 521 (1968).

\bibitem {1990}T. Fukuzawa, E. E. Mendez and J. M. Hong, Phys. Rev.
Lett. \textbf{64}, 3066 (1990).

\bibitem {1994}L. V. Butov \textit{et al.}, Phys. Rev. Lett. \textbf{73}, 304 (1994).

\bibitem {Loz-Ber}Yu. E. Lozovik and O. L. Berman, JETP \textbf{84}, 1027 (1997).

\bibitem {But_ring}L. V. Butov, A. C. Gossard and D. S. Chemla, Nature \textbf{418}, 751 (2002).

\bibitem {snoke_nat}D. Snoke \textit{et al.}, Nature \textbf{418}, 754 (2002).

\bibitem {L1}A. V. Larionov and V. B. Timofeev, JETP Letters
\textbf{73}, 301 (2001).

\bibitem {L2}A. V. Larionov \textit{et al.}, JETP Letters \textbf{75}, 570
(2002).

\bibitem {Snoke2003}D. Snoke \textit{et al.}, Solid State Commun. \textbf{127}, 187 (2003).

\bibitem {MacDonald}J. P. Eisenstein, A. H. MacDonald, Nature \textbf{432}, 691 (2004).

\bibitem {Balats}A. V. Balatsky, Y. N. Joglekar and P. B. Littlewood, Phys. Rev. Lett. \textbf{93}, 266801
(2004).

\bibitem {BLS}L. V. Butov \textit{et al.} Phys. Rev. Lett. \textbf{92}, 117404
(2004).

\bibitem {Rap04}R. Rapaport \textit{et al.}, Phys. Rev. Lett. \textbf{92}, 117405
(2004).

\bibitem {Levitov}L. S. Levitov, B. D. Simons and L. V. Butov, Phys. Rev. Lett. \textbf{94}, 176404
(2005).

\bibitem {Den}S. Denev, S. H. Simon and D. W. Snoke, Solid State Commun. \textbf{134}, 59 (2005).

\bibitem {Ha}M. Haque, Phys. Rev. E \textbf{73} 066207 (2006).

\bibitem {Sug06}A. A. Chernyuk, V. I. Sugakov, Phys. Rev. B \textbf{74}, 085303
(2006).

\bibitem {Szym}M. H. Szymanska, J. Keeling and P. B. Littlewood, Phys. Rev. Lett. \textbf{96}, 230602
(2006).

\bibitem {Tim}V. B. Timofeev and A. V. Gorbunov, J. Appl. Phys.
\textbf{101}, 081708 (2007).

\bibitem {PRB07}Sen Yang \textit{et al.}, Phys. Rev. B \textbf{75}, 033311
(2007).

\bibitem {stern}M. Stern \textit{et al.}, Phys. Rev. Lett. \textbf{101}, 257402 (2008).

\bibitem {polar_nat}J. Kasprzak \textit{et al.}, Nature \textbf{443}, 409
(2006).

\bibitem {But_new}Sen Yang \textit{et al.}, Phys. Rev. B \textbf{81}, 115320
(2010).

\bibitem {PRB10}A. V. Paraskevov and S. E. Savel'ev, Phys. Rev. B \textbf{81}, 193403
(2010).

\bibitem {Deng}H. Deng, H. Haug and Y. Yamamoto, Rev. Mod. Phys. \textbf{82}, 1489
(2010).

\bibitem {com}A. V. Paraskevov, Preprint arXiv:0902.3909 (2009).

\bibitem {Bak}P. Bak, C. Tang and K. Wiesenfeld, Phys. Rev. A \textbf{38}, 364 (1988).

\bibitem {But_rev}L. V. Butov, J. Phys.: Condens. Matter \textbf{16},
R1577 (2004).

\bibitem {cond-mat}L. V. Butov \textit{et al.}, Preprint arXiv:cond-mat/0308117 (2003).

\bibitem {Bak-book}P. Bak, \textit{How nature works: The science of self-organized
criticality} (Springer, New York, 1996).

\bibitem {mob01}H. L. Stormer \textit{et al.}, Phys. Rev. B \textbf{41},
1278 (1990).

\bibitem {mob02}M. P. Lilly \textit{et al.}, Phys. Rev. Lett. \textbf{90}, 056806
(2003).

\bibitem {mob}H. P. van der Meulen \textit{et al.}, Phys. Rev. B \textbf{60},
4897 (1999).

\bibitem {cp}G. R. Facer \textit{et al.}, Phys. Rev. B \textbf{59}, 4622 (1999).

\bibitem {PRB10Er}A. V. Paraskevov and S. E. Savel'ev, Phys. Rev. B \textbf{82}, 119902(E) (2010).

\bibitem {screen}U. Sivan, P.M. Solomon and H. Shtrikman, Phys. Rev. Lett. \textbf{68}, 1196 (1992).

\end{thebibliography}
\end{document}